\documentclass{eas}
\usepackage{graphicx}
\begin{document}

\title{Numerical simulations of Optical Turbulence at low and high horizontal
resolution in Antarctica with a mesoscale meteorological model}
%
\author{F. Lascaux}\address{INAF/Osservatorio Astrofisico di Arcetri,Largo E. Fermi 5, 50125 Florence, Italy}
\author{E. Masciadri}\sameaddress{1}
\author{S. Hagelin}\sameaddress{1}
\author{J. Stoesz}\sameaddress{1}
\begin{abstract}
It has already been demonstrated that a mesoscale meteorological model such as
Meso-NH (Lafore \etal~\cite{laf1998}) is highly reliable in reproducing 3D maps of
optical turbulence (Masciadri \etal~\cite{mas1999}, Masciadri and Jabouille ~\cite{mas2001},
Masciadri \etal~\cite{mas2004}).
Preliminary measurements above the Antarctic Plateau have so far indicated a pretty good value for the
seeing: 0.27" (Lawrence \etal~\cite{law2004} ), 0.36" (Agabi \etal~\cite{aga2006}) or 0.4"
(Trinquet \etal~\cite{tri2008}) at Dome C. However some uncertainties remain.
That's why our group is focusing on a detailed study of the atmospheric
flow and turbulence in the internal Antarctic Plateau.
Our intention is to use the Meso-NH model to do predictions of the
atmospheric flow and the corresponding optical turbulence in the internal
plateau. The use of this model has another huge advantage: we have access to
informations inside an entire 3D volume which is not the case with
observations only.
Two different configurations have been used: a low horizontal resolution (with
a mesh-size of 100 km) and a high horizontal resolution with the grid-nesting
interactive technique (with a mesh-size of 1 km in the innermost domain
centered above the area of interest).
We present here the turbulence distribution reconstructed by Meso-NH for 16
nights monitored in winter time 2005, looking at the the seeing
and the surface layer thickness.
\end{abstract}
\maketitle
\section{Introduction}
The extreme low temperatures, the dryness, the typical high altitude of the
internal Antarctic Plateau (more than 2500~m), joint to the fact that the optical
turbulence seems to be concentrated in a thin surface layer whose thickness is of
the order of a few tens of meters do of this site a place in which, potentially,
we could achieve astronomical observations otherwise possible only by space.
Despite exciting first results (Lawrence \etal~\cite{law2004}; Agabi \etal~\cite{aga2006}; 
Trinquet \etal~\cite{tri2008}) making the internal Antarctic Plateau a site of potential great interest for astronomical applications, some uncertainties still remain.
Here we studied the Dome C area with a mesoscale meteorological model (Meso-NH, Lafore \etal~\cite{laf1998}).
Numerical simulations offer the advantage to provide volumetric maps of the optical
turbulence ($C_N^2$) extended on the whole internal plateau and, ideally, to
retrieve comparative estimates in a relative short time and homogeneous way on
different places of the plateau.
Fifteen winter nights (the same as from Trinquet \etal~(\cite{tri2008}) were simulated.
Using the forecasted $C_N^2$ profiles, we retrieved the surface layer thicknesses H$_{SL}$ and the free atmosphere 
seeing ($\epsilon{_{FA}}$) for all 15 nights.
\par
This study is a short survey of the more detailed study available in Lascaux \etal~(\cite{las2009}).

\section{Surface layer seeing and free atmosphere seeing}
Two different configuration of Meso-NH were chosen: a low horizontal resolution mode, and a high horizontal resolution 
mode (with the grid-nesting interactive technique). 
To know more about the numerical set-up and the model configuration, the reader can 
refer to Lascaux \etal~(\cite{las2009}).
In that paper can also be found a validation of the model with comparisons of meteorological parameters (wind and 
temperature) at Dome C between model outputs and observations.
One of the conclusion is that both configurations generated better forecast for wind speed and temperature than 
the analysis from the ECMWF, especially near the surface. More over, the grid nested mode gave better results than the 
low resolution mode.
\par
In order to verify how well the simulated H$_{SL}$ matches with the measured one we 
computed the typical height of the surface layer (averaged each night between 12 UTC and 16 UTC) 
using the same criterion as in Trinquet \etal~(\cite{tri2008}):
\begin{equation}
 \label{eq:bl1}
 \frac{ \int_{8m}^{h_{sl}} C_N^2(h)dh }{ \int_{8m}^{1km} C_N^2(h)dh } < 0.90
\end{equation}
where $C_N^2$ is the refractive index structure parameter.
\par
\begin{figure}
\begin{center}
\includegraphics[width=0.4\textwidth]{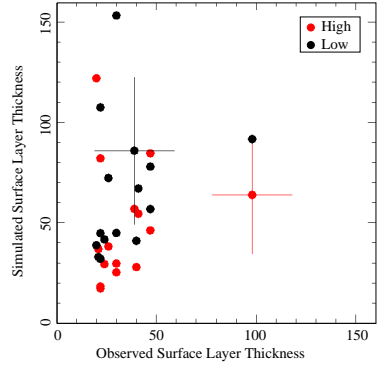}
\end{center}
\caption{Correlation plot between measured and simulated surface layer thicknesses (black: monomodel
configuration; red: grid-nested configuration). For the simulated values only the mean values between 12 UTC
and 16 UTC are considered.
For each configuration of the simulation (high and low horizontal resolution) the error bars are
reported for one point only (and are equal to $\sigma$). Units are in meter (m).}
{\label{fig_corr1}}
\end{figure}
\begin{figure}
\begin{center}
\includegraphics[width=0.8\textwidth]{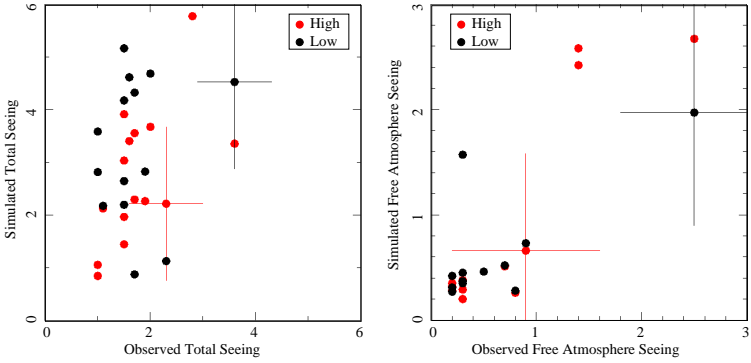}
\end{center}
\caption{Correlation plot between measured and simulated total (top) and free atmosphere (bottom) seeing
(black: monomodel configuration; red: grid-nested configuration).
For the simulated values only the mean values between 12 UTC and 16 UTC are considered.
For each configuration of the simulation (high and low horizontal resolution) the error bars are
reported for one point only (and are equal to $\sigma$). Units are in arcsec.}
{\label{fig_corr2}}
\end{figure}

The observed mean H$_{SL}$ for the 15 winter nights was of 35.3 $\pm$ 5.1~m. The low horizontal resolution 
mode gave a result almost twice higher: H$_{SL,LOW}$=65.9$\pm$ 8.7~m. 
The grid-nesting mode gave better results, comparable to the observations: H$_{SL,HIGH}$=48.9 $\pm$ 7.6~m.
Using these computed mean H$_{SL}$, we deduced the median free atmosphere seeing using the same method as 
in Trinquet \etal~(\cite{tri2008}).
Using H$_{SL,OBS}$=30~m (computed on a larger sample), Trinquet \etal~(\cite{tri2008}) found 
$\epsilon{_{FA,OBS}}$=0.3 $\pm$ 0.2~arcsec.
For the low resolution mode (H$_{SL,LOW}$=65.9~m), the corresponding free atmosphere seeing is slightly overestimated: 
$\epsilon{_{FA,LOW}}$=0.42$\pm$ 0.28~arcsec.
However, the grid-nested mode (H$_{SL,HIGH}$=48.9~m) gave excellent result: $\epsilon{_{FA,HIGH}}$=0.35 $\pm$ 
0.24~arcsec, thus confirming the importance of the high horizontal resolution configuration to obtain reliable 
forecasts.
The corresponding correlation plots (for all 15 nights) are displayed on Figures \ref{fig_corr1} (surface layer 
thickness) and \ref{fig_corr2} (free atmosphere and total seeings).

\par
\begin{figure}
\begin{center}
\includegraphics[width=\textwidth]{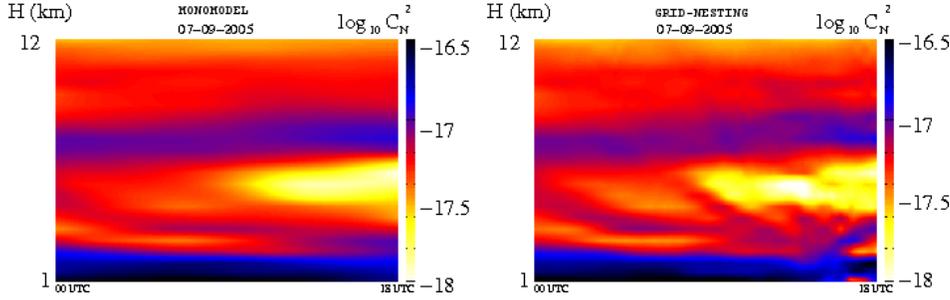}
\end{center}
\caption{Temporal evolution on 18 hours of the C$_N^2$ profiles in the vertical slab (1-12) km related to one winter 
night (04 July 2005), in logarithmic scale. On the left, in the low horizontal resolution mode; on the right, in the 
high horizontal resolution model.}
{\label{fig3}}
\end{figure}

On Fig. \ref{fig3} is displayed the temporal evolution of the C$_N^2$ profile in the free atmosphere (1-12 km vertical 
slab) for one night. 
This night is a good example of how the model is active even at such an altitude.
The vertical distribution of the optical turbulence changes in time with a non negligible dynamic from a quantitative 
point of view. It is even more visible in the high horizontal resolution mode (grid-nesting). 

\section{Conclusion}
We studied the performances of the Meso-Nh mesoscale model in reconstructing optical turbulence profiles, looking at 
the Dome C area, in the internal Antarctic Plateau.
This study was focused on the winter season.
The results concerning the optical turbulence computations are resolution dependent.
A high horizontal resolution mode seems to be mandatory to realize realistic optical turbulence forecast.
In high horizontal resolution mode, Meso-Nh gave excellent results: H$_{SL,HIGH}$=48.9$\pm$ 7.6~m, to be compared to
H$_{SL,OBS}$=35.3 $\pm$ 5.1~m.
The resulting free atmosphere seeing is $\epsilon{_{FA,HIGH}}$=0.35 $\pm$ 0.24~arcsec, very close to the observed one, 
$\epsilon{_{FA,OBS}}$=0.3 $\pm$ 0.2~arcsec.

\section*{Acknowledgements}
This study has been funded by the Marie Curie Excellence Grant (FOROT) - MEXT-CT-2005-023878.

\end{document}